# Patched RTC: evaluating LLMs for diverse software development tasks

Asankhaya Sharma, Patched Codes, Inc
asankhaya@patchedcodes.com

## Abstract

This paper introduces Patched Round-Trip Correctness (Patched RTC), a novel evaluation technique for Large Language Models (LLMs) applied to diverse software development tasks, particularly focusing on "outer loop" activities such as bug fixing, code review, and documentation updates. Patched RTC extends the original Round-Trip Correctness method to work with any LLM and downstream task, offering a self-evaluating framework that measures consistency and robustness of model responses without human intervention. The study demonstrates a correlation between Patched RTC scores and task-specific accuracy metrics, presenting it as an alternative to the LLM-as-Judge paradigm for open-domain task evaluation. We implement Patched RTC in an open-source framework called patchwork, allowing for transparent evaluation during inference across various patchflows. Experiments comparing GPT-3.5 and GPT-4 models across different software development tasks reveal that Patched RTC effectively distinguishes model performance and task difficulty. The paper also explores the impact of consistency prompts on improving model accuracy, suggesting that Patched RTC can guide prompt refinement and model selection for complex software development workflows. Our implementation is open-source and available at https://github.com/codelion/optillm

# Introduction

In the past couple of years, LLMs have shown great progress in helping developers with various software development tasks. Typical evaluation of LLMs on coding related tasks focuses mostly on “first-party” (or inner development loop) problems like code generation, summarization and unit testing. Most of such tasks happen within the IDE of the developer, often assisted by a GitHub Copilot-like plugin. Relatively little attention has been paid to the “second-party” (or outer development loop) tasks like bug fixing, code review, refactoring, pull requests, code integration, documentation updates and security patching. We argue that a large majority of software development time is spent in these second-party outer loop activities v/s actual coding. Accelerating software development requires us to automate these tasks and LLMs can be used to do that effectively.

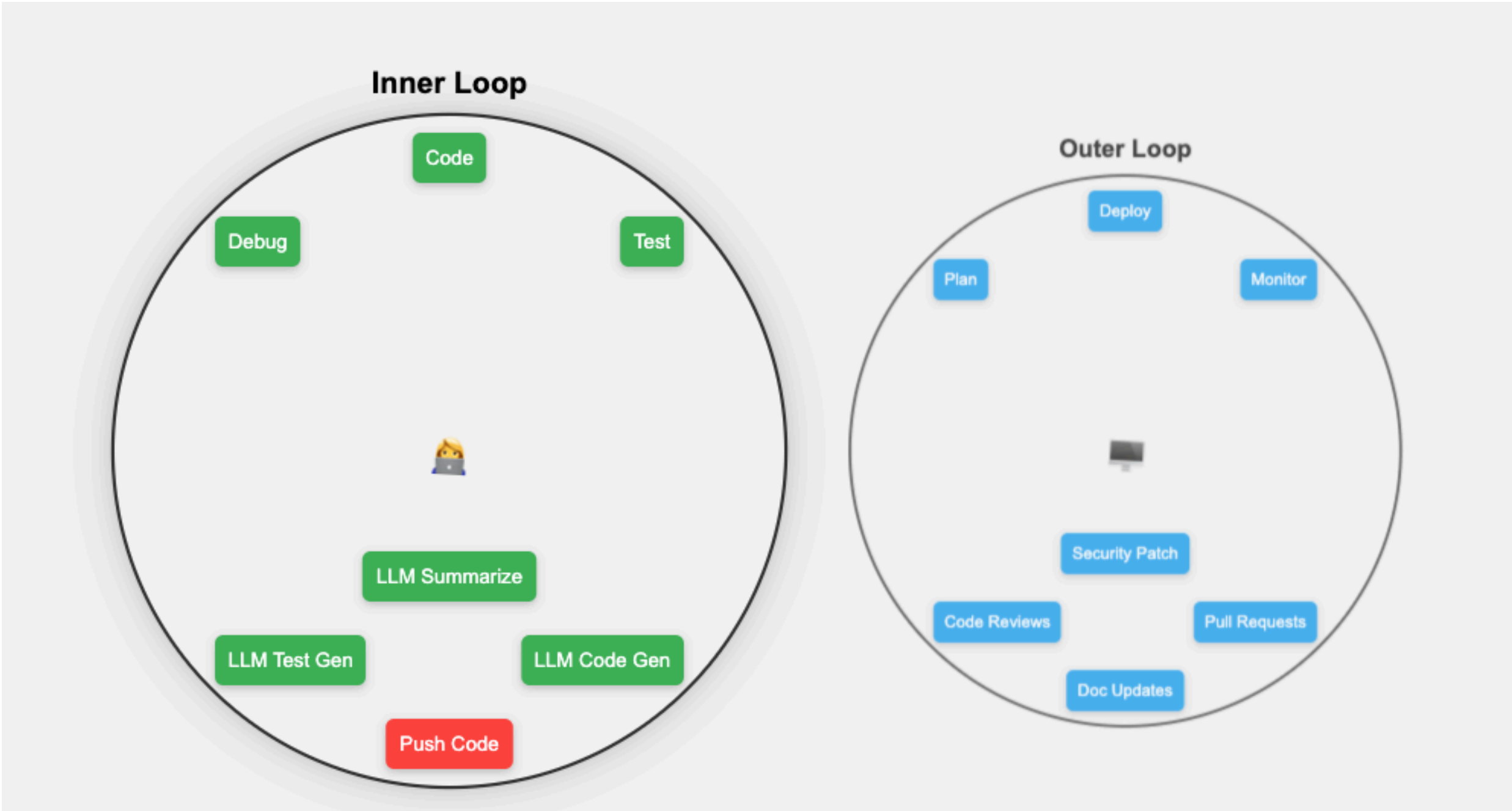

[Inner Loop v/s Outer Loop Visualization]

In order to ascertain the effectiveness of LLMs when it comes to automating developer outer-loop tasks we need a mechanism for good evaluation. The most popular benchmark to evaluate LLMs on coding related tasks are HumanEval [2] (and MBPP) and its subsequent extensions like HumanEvalPack and EvalPlus. The benefit of these benchmarks is that they can be run completely unsupervised and the evaluation of results does not require any human intervention or review. However, these benchmarks do not adequately capture real-world scenarios. Though there have been attempts by other complex benchmarks like the bigcodebench or task-specific ones like static-analysis-eval, the current gold standard in LLM evaluation is the LMSYS Chatbot Arena, where humans rate model responses via pairwise comparison for an elo rating system.

As of the date of publishing this post, the coding category on the Arena is led by the frontier models from Anthropic, OpenAI and Google:

Category: Coding

**Coding: whether conversation contains code snippets**

**#models: 109 (95%) #votes: 277,419 (19%)**

| Rank* (UB) | Delta | Model | Arena Score | 95% CI | Votes | Organization |
|---|---|---|---|---|---|---|
| 1 ↑ | 1 | Claude 3.5 Sonnet | 1302 | +8/-11 | 5081 | Anthropic |
| 1 | 0 | GPT-4o-2024-05-13 | 1297 | +7/-8 | 12379 | OpenAI |
| 3 | 0 | Gemini-1.5-Pro-API-0514 | 1266 | +8/-8 | 10284 | Google |

Evaluating models on the Arena is expensive and time-consuming as it requires crowd sourced inputs from humans. To combine the best of both worlds and based on the experience of Chatbot Arena, newer unsupervised evaluation benchmarks have been proposed (e.g. Arena-Hard-Auto) that show a high correlation with the original human-rated results reported in the arena. These benchmarks use the LLM-as-Judge (or Jury) paradigm [3] and generate the scores automatically without human review. Any concerns around contamination of benchmark data can be addressed via private or continuously updating datasets like the livecodebench and livebench.

In this work, we propose a technique for model evaluation (Patched RTC) that is based on the notion of Round-trip Correctness (or RTC). This approach was first introduced by Google Deepmind and applied to code LLMs [1]. We extend and expand the original technique to work for any LLM and any downstream task. In particular, the key contributions of Patched RTC are:

- It is an evaluation technique that is generic and works with all LLMs.
- It can be applied transparently during inference to self-evaluate the responses by the model without requiring any code changes.
- It can be applied to a wide domain of tasks and applications where the evaluation of correctness is difficult due to a lack of human annotations.
- It works extremely well for outer-loop software development tasks (like bug fixing, pull request reviews, and documentation updates) where we are working with patches (or commits) instead of code.

# Approach

The generic implementation of Patched RTC is simple and works as follows:

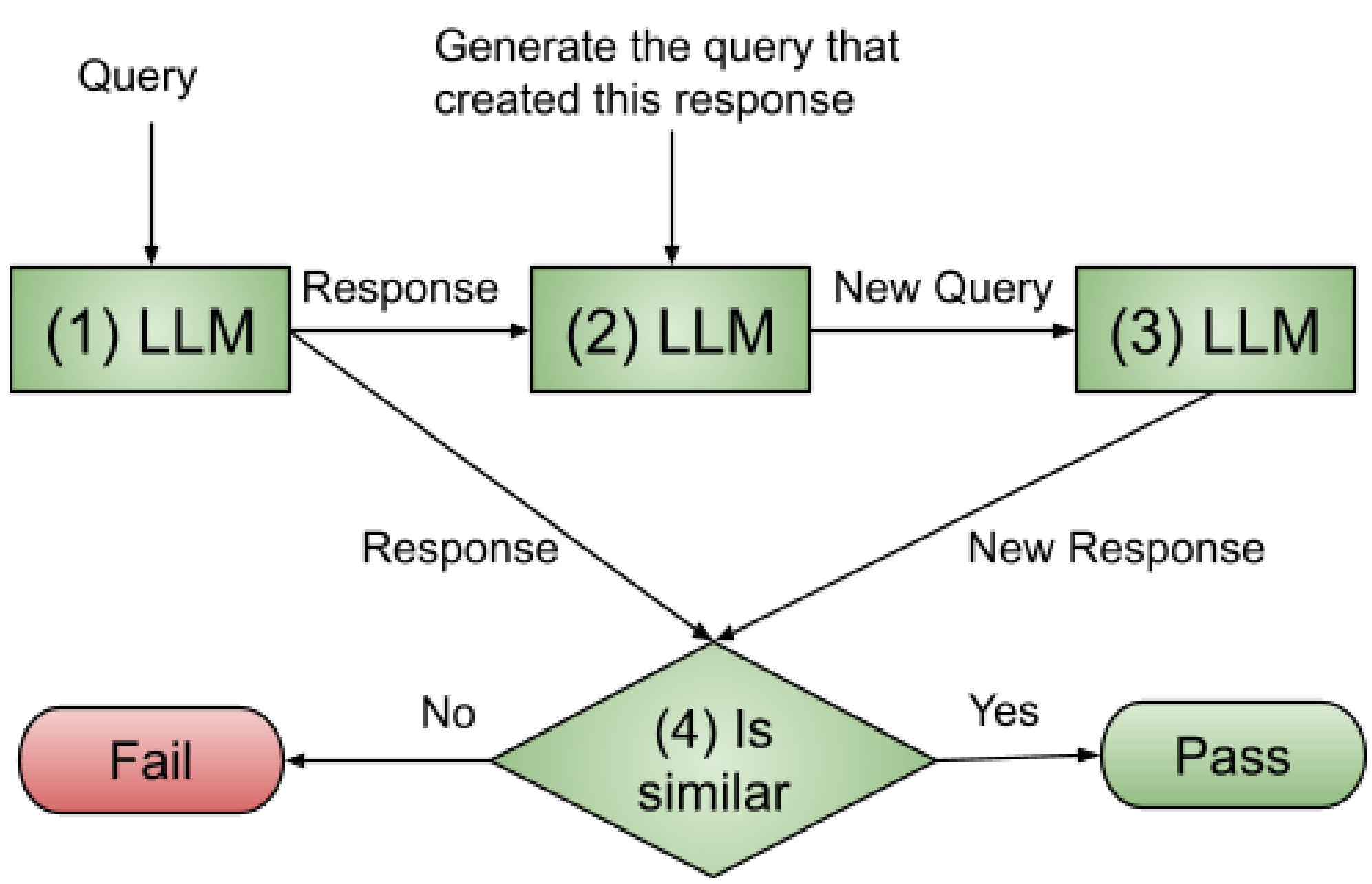

Say we have the model M that is used to generate a response R for the user query Q. Now, we wish to evaluate if the response R for the Q is "correct".

1) Q → [M] → R

We take Q, R and prompt the model to generate an alternate query Q1 such that Q1 is sufficient to recreate the response R.

2) Q, R → [M] → Q1

Now, we take the new query Q1 and ask the model to generate another response R1.

3) Q1→ [M] → R1

Finally, we check if R and R1 are similar by computing a similarity score (0-1).

4) R, R1 → [M] → score

If score > threshold (say 0.8), we say that response R (for the query Q) is correct (w.r.t. RTC).

Step 4) can also be done without the use of LLMs, if we choose to rely on another similarity metric like cosine similarity (or number of unit tests passed in case of code generation).

# Patchflows

We define patchflows as workflows that automate outer-loop development tasks like bug fixes, pull request reviews, documentation updates and library upgrades. Our open-source framework patchwork makes it easy for developers to build and run patchflows. One of the challenges with using LLM-assisted patchflows is that it is hard to evaluate the effectiveness of using them in practice.

Patched RTC can be easily adopted to evaluate patchflows as follows:

A patch (or commit) has two parts before_code and after_code.

Patchflows either have 1) a patch as input (e.g. for pull request review) or 2) generate a patch as an output (e.g. bug fixes). For the user prompt Q and response R we can handle these two cases as:

1) Q, before_code, after_code → [M] → R

   Applying Patched RTC, we first generate alternate query Q1

Q, before_code, after_code, R → [M] → Q1

Q1, before_code, after_code → [M] → R1

R, R1 → [M] → score

If score > threshold, R is correct (w.r.t RTC).

2) Q, before_code → after_code

Similar to above, we first generate alternate query Q1

Q, before_code, after_code → [M] → Q1

Q1, before_code → after_code1

Since both after_code and after_code1 are code, we can actually use a stronger measure of similarity. We use exact match as the notion of similarity thus,

If exact_match(after_code, after_code1), R is correct (w.r.t. RTC).

Almost all patchflows and the corresponding tasks can be classified in either one or the other category. We list some of these tasks in the table below:

| 1) Tasks taking a patch as input | 2) Tasks producing a patch as output |
|---|---|
| Code Review | Bug fixing |
| Code style and convention checking | Code refactoring |
| Performance impact analysis | Code optimization |
| Dependency update impact analysis | Documentation generation and updates |
| Test coverage analysis | Test case generation |
| Documentation consistency check | Code comment generation or improvement |
| Changelog generation | Automated code formatting |
| Release notes compilation | Internationalization and localization updates |
| Commit message quality assessment | Configuration file updates |
| Code complexity analysis | Automated code migrations |
| Refactoring suggestions | Dependency version updates |
| Merge conflict detection and resolution | API documentation updates |
| License compliance checking | Database schema migrations |

Without using an unsupervised technique like Patched RTC, it would be really hard to evaluate the correctness of LLMs when applied for such tasks as it would require the presence of human annotations or checks for each of these tasks. We have implemented several such tasks as patchflows in our open-source framework patchwork:

1. **AutoFix**: Generate and apply fixes to code vulnerabilities in a repository.
2. **PRReview**: On PR creation, extract code diff, summarize changes, and comment on PR.
3. **GenerateDocstring**: Generate docstrings for methods in your code.
4. **GenerateREADME**: Create a README markdown file for a given folder, to add documentation to your repository.
5. **ResolveIssue**: Identify the files in your repository that need to be updated to resolve an issue (or bug) and create a PR to fix it.

One of the common challenges in adoption of these patchflows by developers is the assurance around accuracy and consistency of the outputs. In the next section, we will see how we can use Patched RTC to address this issue.

# Evaluation

We first demonstrate the usefulness of Patched RTC across a generic set of diverse tasks by comparing it with the Arena-Hard-Auto benchmark. The below table shows the performance of different models when evaluated with RTC v/s the LLM-as-Judge paradigm as is standard in Arena-Hard-Auto. We run our tests at a high similarity threshold (0.95).

As seen from the table below, we notice that is a correlation (with Pearson coefficient of 0.81) when compared to the numbers in Arena-Hard-Auto, thus showing that Patched RTC can be used as an evaluation mechanism instead of LLM-as-Judge for generic and diverse tasks. However, there are some differences when compared to Arena-Hard-Auto as well. We have gpt-4-0125-preview as the performing best model on Patched RTC and llama-3-70b-instruct also performs better than gpt-4o. These differences arise because Patched RTC measures robustness and consistency by checking the model's ability to invert itself and that may not necessarily be the same as alignment with desired responses (as rated by humans).

| Model | Patched RTC | Arena-Hard-Auto |
|---|---|---|
| gpt-4-0125-preview | 76.6 | 78.0 |
| llama-3-70b-instruct | 47.4 | 46.6 |
| gpt-4o | 44.2 | 79.2 |
| llama-3-8b-instruct | 28.4 | 20.6 |
| gpt-3.5-turbo-0125 | 22.6 | 23.3 |

Next, we apply Patched RTC to compare the performance of different patchflows. The following table shows the numbers for each of the patchflows supported by our open-source framework [patchwork](). We selected a sample of the most active GitHub repositories in 3 different languages (Python, Java and JavaScript). Then we ran the patchflows on these repositories including their issues and pull requests on the main branch. We ran each patchflow only once; however a patchflow may make several calls to the LLM during the run depending on how it is implemented.

We ran these experiments at a similarity threshold of 0.8 as we found that higher thresholds tend to reject many responses that are equivalent due to small changes in either the comment or structure of the generated code. We chose to compare between gpt-3.5-turbo and gpt-4o models for these experiments as they are the most used models with our framework based on usage data. These two models also provide excellent trade-offs in price v/s performance. We expect the results to generalize to other models. The next chart shows the performance when comparing these models with Patched RTC.

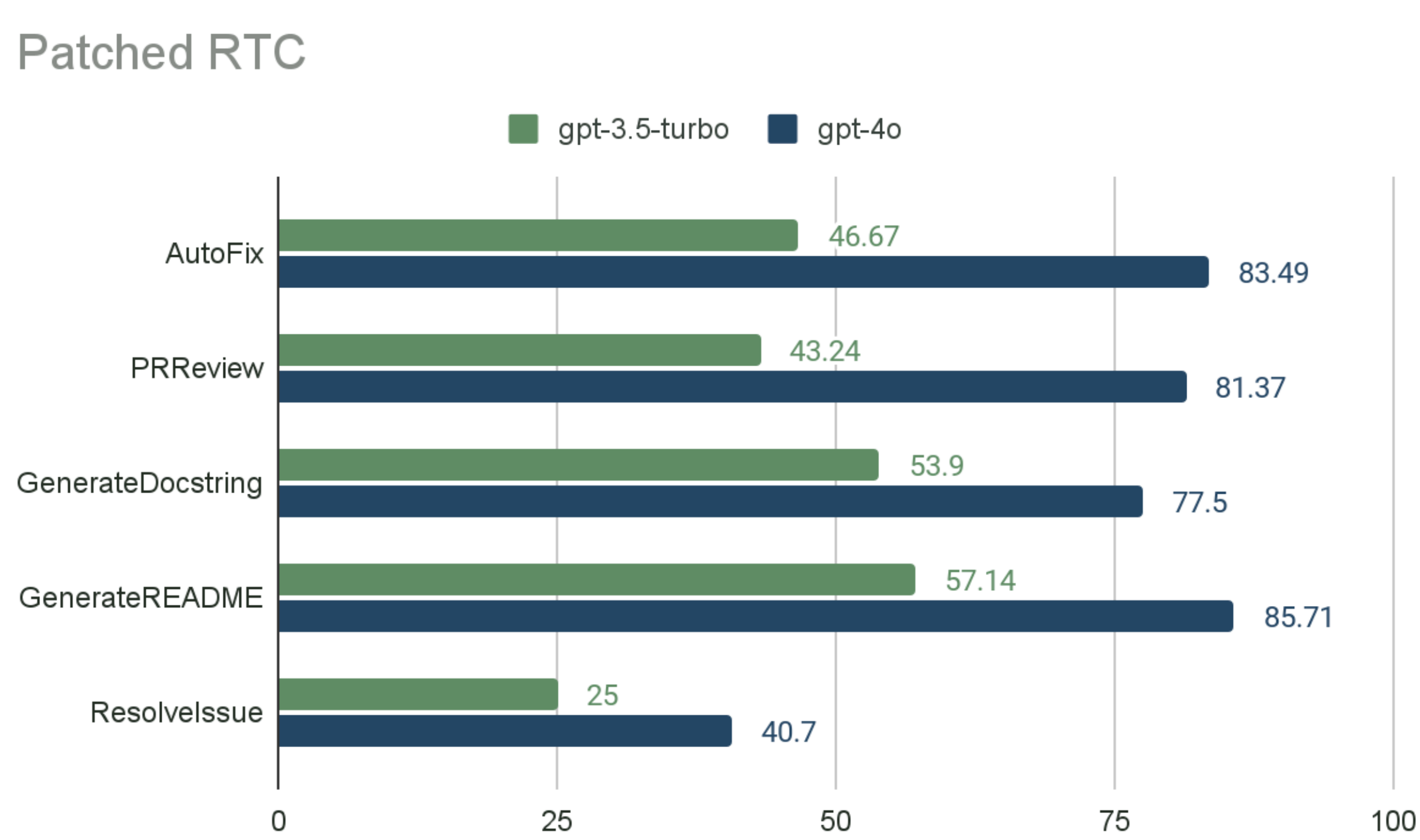

Unsurprisingly, we can see that gpt-4o performs better than gpt-3.5-turbo across all the tasks but for some of the more complex patchflows like AutoFix and PRReview the difference between the two models is more pronounced. This suggests that a model with better reasoning capabilities like gpt-4o is needed for a patchflow like AutoFix and using a less capable model will not be sufficient.

Another thing we can notice from above is that certain tasks are just harder than others, we see that ResolveIssue is the most different patchflow as both models have the lowest RTC pass scores on that. On the other hand, we see that GenerateREADME is one of the easier tasks as both models' scores are highest on that task. Patched RTC is useful to compare model performance across diverse tasks.

Now, in order to check if better performance on Patched RTC does indeed correlate with actual accuracy on the task we need to evaluate the responses further. Usually, this is the hardest part of designing an eval for a new task. In absence of expensive human annotation and reviews we can define oracles to make the final judgment of accuracy. Oracles are task specific and need to be carefully designed to ensure they capture the intended definition of accuracy.

For instance, in the AutoFix patchflow we can use a static analyzer (Semgrep) as an oracle. We scan the fixed code with Semgrep to ascertain if the vulnerability has indeed been fixed [4]. (Similarly for the ResolveIssue patchflow unit test results can serve as an oracle.) The below table shows the results when using a static analyzer as an oracle:

| Tests | Vulns | Original Fix % | RTC Pass % | Fix % | RTC Fail % | Fix % |
|---|---|---|---|---|---|---|
| 103 | 106 | 52.8 | 83.5 | 55.2 | 16.5 | 42.1 |

In total there are 103 tests in the AutoFix dataset which correspond to 106 vulnerabilities (there may be more than 1 instance of a vulnerability in a test). We define the Fix % as the percentage of vulnerabilities that are fixed by the AutoFix patchflow. It is calculated as follows:

$$\text{Fix \%} = \frac{\text{(No of vulns before running AutoFix - No of vulns after running AutoFix)}}{\text{No of vuln before running AutoFix}}$$

Based on the results, we see that the actual fix rate (or accuracy) on the AutoFix task is 52.8 v/s the RTC Pass score which was 83.5. But we see that the fix rate for responses that pass Patched RTC is higher (55.2) v/s those that fail (42.1). This suggests that RTC is able to distinguish more accurate responses by measuring robustness (or consistency [5]). To test this hypothesis we add a very simple one-line consistency prompt (this is very similar to the “think step-by-step” prompt that seems to help models do better reasoning) to all the tests and check if this improves the fix rate.

Consistency prompt:
*"Respond with clarity, consistency, and precision, maintaining a structured format throughout.”*

The above prompt was prepended to the system prompt of the request for all the tests and we computed the Fix % of the responses. We saw that this improves the Fix rate by 14.4%.

| Tests | Vulns | Original Fix % | Consistency Prompt Fix % |
|---|---|---|---|
| 103 | 106 | 52.8 | 60.4 |

This shows that making responses more consistent can improve accuracy and Patched RTC pass rate can be used as an indicator of how well the model will perform on the task. Next, we use the consistency prompt and reevaluate the Patched RTC pass rate for all the tasks. We see that adding this prompt in general improves the overall pass rate across the patchflows for both the models - gpt-3.5-turbo and gpt-4o. The increase is more pronounced and consistent in a more capable model like gpt-4o v/s gpt-3.5-turbo as can be seen in the radar charts below.

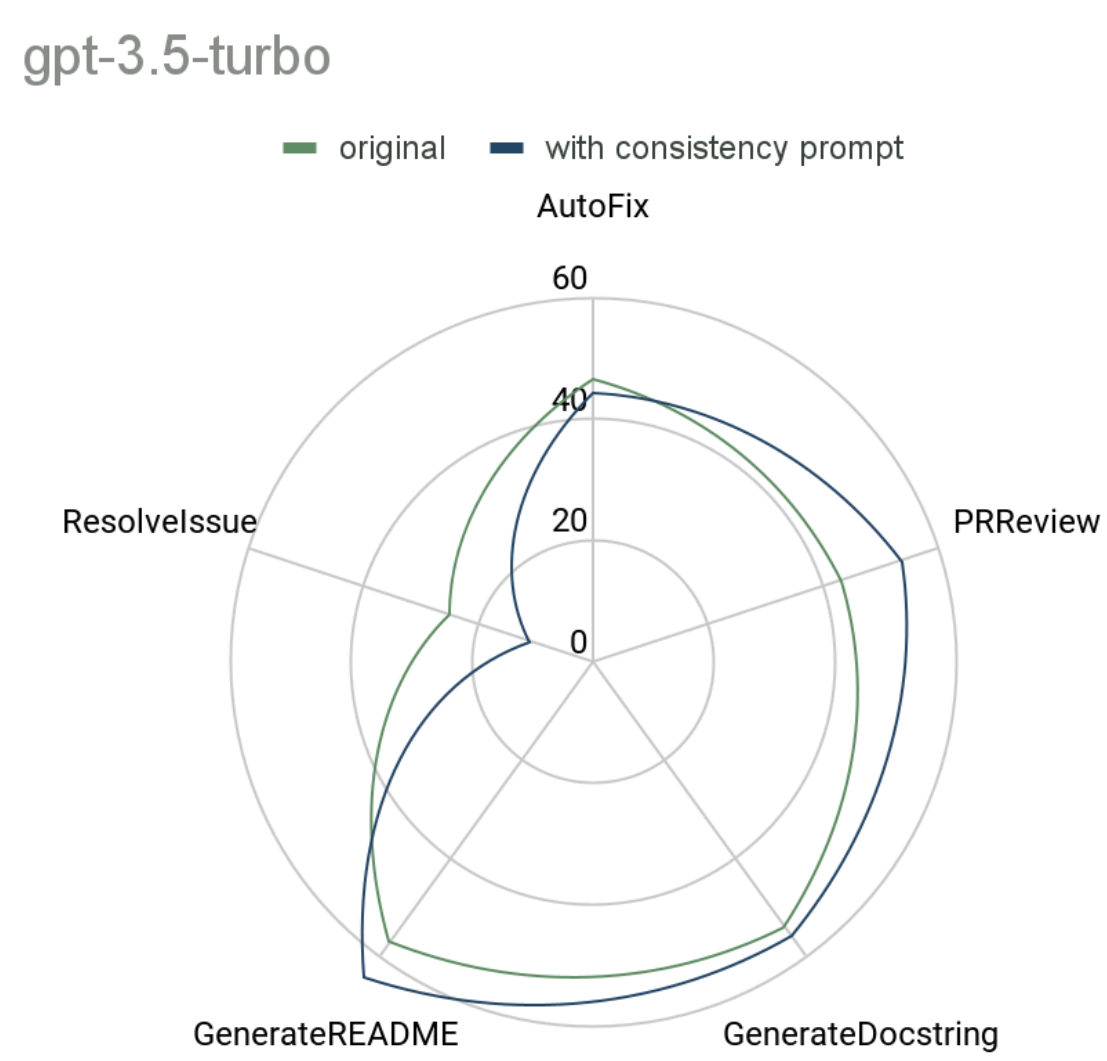

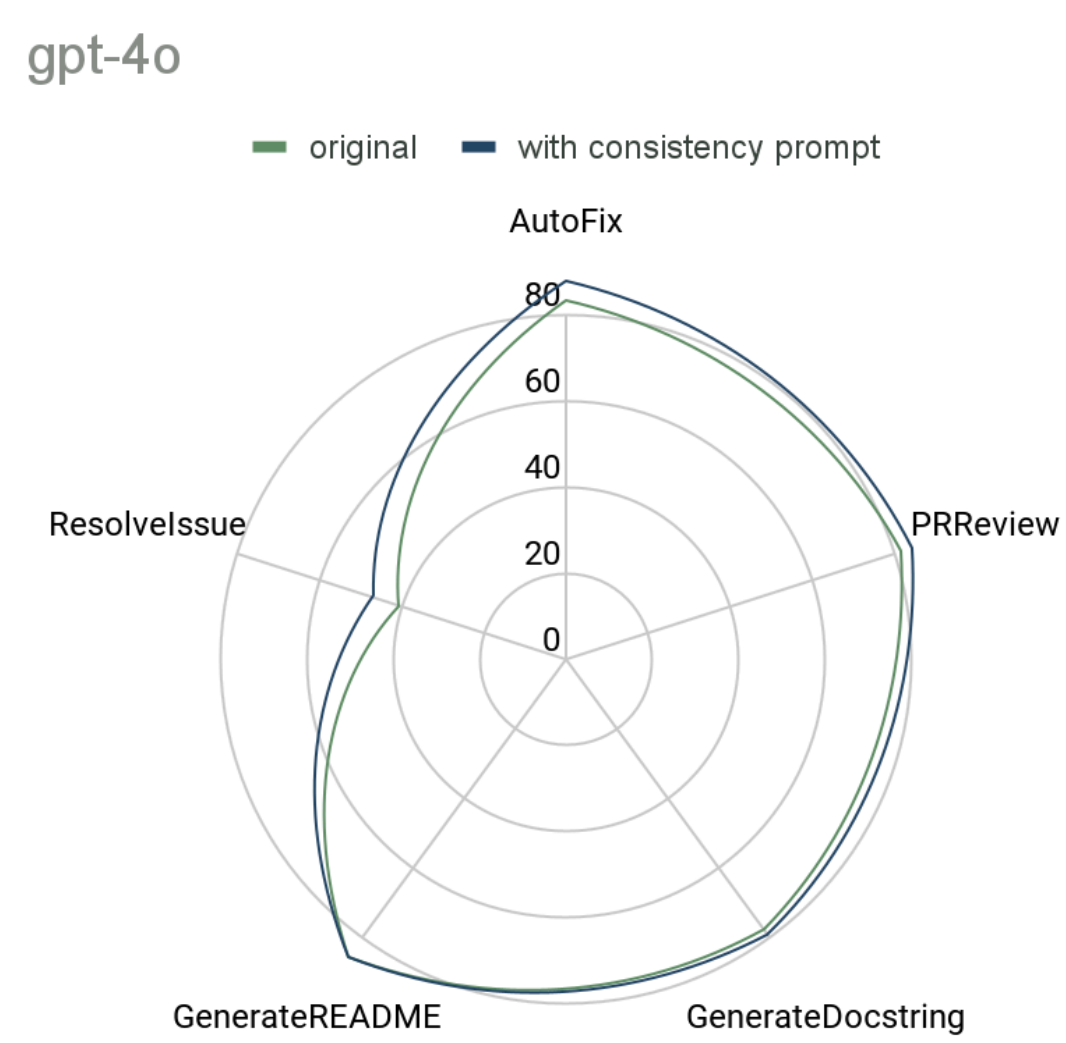

# Discussion

The use of Patched RTC does incur an additional inference cost of ~3x depending on how the similarity measure is computed. Thus, it is more likely to be useful during testing and evaluation of patchflows and guiding the refinement of the prompts. When used as an active inference technique the improvements in accuracy and robustness need to be balanced with the increased inference cost. Also, the similarity threshold and similarity measure are likely to be dependent on the task and experimenting with a few options before choosing one will likely lead to better results.

That said, we have actually found it quite useful for evaluating open-domain tasks in software development as it is hard to ascertain the accuracy of models or complex workflows on these tasks without human annotations or reviews. We have also used Patched RTC to evaluate different inference time scaling techniques like best-of-n and mixture of agents [7].

## What does Patched RTC measure?

RTC isn't strictly measuring "correctness" in the traditional sense, as we don't have a ground truth to compare against. Instead, it's measuring something more nuanced:

1. Consistency: RTC evaluates how consistently the model can reproduce similar content given a description of its own output.
2. Robustness: It tests whether the model's output is stable enough that it can be approximately reproduced from a summary of itself.
3. Coherence: It checks if the model's output contains enough clear, structured information that another instance of the model can grasp and reproduce the key points.
4. Self-invertibility: It measures how well the model can "invert" its own output - turning a response into a query and back into a similar response.

The key benefits of Patched RTC are:

- It is a different form of evaluation as it tests the ability of the LLM to act as an invertible function.
- It does not require the use of Judge (or Jury) LLMs and can be done with a single model.
- It is an unsupervised evaluation as we do not rely on any human annotation or checks.
- It can be used along with any existing benchmark to see how Patched RTC correlates with them.
- It can be used for a wider spectrum of domains that do not have good human annotations.

Our work on RTC is just a beginning, there are a lot of directions we can explore further:

- Using a different model for round-trip response compared to the original model.
- Optimize the prompts automatically to generate more consistent responses.
- Impact of different oracles on task accuracy when used with Patched RTC.

# Conclusions

In this article, we introduced Patched RTC, a self evaluating framework that works across diverse tasks. Patched RTC measures consistency and robustness of LLM responses and is correlated with oracle based accuracy metrics. It presents an alternative to the LLM-as-Judge paradigm that is currently one of the most common ways to evaluate models for open-domain tasks. We also showed that making prompt changes that increase consistent responses from models do help in improving the overall accuracy of the model.

Our implementation is open-source [6] and is also integrated in patchwork, anyone building patchflows can make use of RTC to evaluate and optimize it for their own downstream task.

## Usage

To get access to Patched RTC:

Use the `patched_api_key` with our OpenAI compatible endpoint available at patched.codes and just change the base url to `https://patchwork.patched.codes/evaluate/v1`. When using this endpoint only those responses that pass Patched RTC will be generated, otherwise the response will be empty. If you want to compare with how the response would have been without Patched RTC, you can send the same request through our usual OpenAI compatible endpoint at `https://patchwork.patched.codes/v1`.